\documentstyle[aps,prl,epsfig,tabularx]{revtex}

\def\thebibliography#1{\section*{References\markboth
  {REFERENCES}{REFERENCES}}\list
  {[\arabic{enumi}]}{\settowidth\labelwidth{[#1]}\leftmargin\labelwidth
  \advance\leftmargin\labelsep
  \usecounter{enumi}}
  \def\newblock{\hskip .11em plus .33em minus -.07em}
  \sloppy
  \sfcode`\.=1000\relax}
\newcommand{\sect}{\section}
\begin{document}
\renewcommand{\sect}[1]{\section{#1}\setcounter{equation}{0} }

\newcommand{\eqr}{\begin{eqnarray}}
\newcommand{\rqe}{\end{eqnarray}}

\newcommand{\eq}{\begin{equation}}
\newcommand{\qe}{\end{equation}}


\title{Empirical investigation of a quantum field theory of forward rates}

\author{Belal E. Baaquie and Srikant Marakani}
\address{Permanent address: Department of Physics,
National University of Singapore, Kent Ridge Road, 
Singapore - 091174}

\maketitle

\begin{abstract}
  
  A new test of a wide class of interest rate models is proposed and
  applied to a recently developed quantum field theoretic model and
  the industry standard Heath-Jarrow-Morton model. This test is
  independent of the volatility function unlike other tests previously
  proposed in the literature. It is found that the HJM model is
  inconsistent with the data while the quantum field theoretic model
  is in significant agreement with data. We also show that a portion
  of the spread between long and short term interest rates is
  explicable in terms of this model. 
\end{abstract}




Physicists have been working on several aspects of financial research
over the last decade \cite{bouchcont}. One of the most important and as yet unsolved
problems in finance is the modelling of interest rates. The interest
rates at any point in time form a usually continuous curve (current
interest rates for different times in the future) called the forward
rate curve (FRC). We denote these rates by $f(t, x)$ where $t$ is the
current time and $x$ is the time in the future for which the forward
rate applies. For example, $f(1, 2)$ is the interest rate one year
from now for an instantaneous deposit to be made 2 years into the future.

The earliest interest rate models (eg., Vasicek \cite{vasicek}) dealt
only with the spot rate (current interest rate for the present time)
and the forward rate curve was treated as a derived quantity. These
were found to be inconsistent with the observed data. The Ho-Lee
\cite{holee} and HJM\cite{hjm} models were developed to deal with this
problem by modelling the entire FRC rather than just the spot rate.
The HJM model is, however, limited in the sense that the Brownian
motions on which the HJM economy depends are independent of $x$. One
way of removing this restriction is by formulating the forward rate
curve as a quantum mechanical string as in Baaquie \cite{baaquie} (we
will refer to this model as the quantum field theoretic model).

Several empirical tests of the HJM model have been performed (eg.,
B\"uhler, Uhrig-Homburg, Walter and Weber \cite{buhler}, Flesker
\cite{flesker}, Sim and Thurston \cite{sim}) with mixed results. All
of the tests assume a certain form for the volatility function
$\sigma$. In this paper, we propose a test which is {\em independent}
of the volatility function. The test is applied to the HJM and quantum
field theoretic model which was introduced in \cite{baaquie}. The HJM
model can be formulated as a limit of the model in \cite{baaquie}
which is briefly reviewed below.

The one factor quantum field theoretic model models the forward rates
as 
\begin{equation} 
\label{qft}
\frac{\partial f(t, x)}{\partial t} = \alpha (t, x) + \sigma (t, x) A (t, x)
\qe
where $A(t, x)$ is a quantum field whose action is given by
\begin{eqnarray}
\label{sf}
S[A]&=&\int_{t_0}^{\infty}dt\int_t^{t+T_{FR}}dx{\cal L}[A]\\
\label{lf}
{\cal L}[A]&=&-\frac{1}{2}\left(A^2(t,x) +
\frac{1}{\mu^2}\left(\frac{\partial A(t, x)}{\partial x}\right)^2
\right)\nonumber\\ 
\end{eqnarray} 
where $T_{FR}$ is the largest time to maturity for which the forward
rates are defined (the domain is hence a semi-infinite parallelogram
defined by $t>t_0, t<x<t+T_{FR}$). $T_{FR}$ is introduced to
ensure that the action is well defined but does not affect final
results as the limit $T_{FR} \rightarrow \infty$ must be taken.
When $\mu \rightarrow 0$, this model reduces to the HJM model upto a
rescaling (for details, please see \cite{baaquie}).

We assume that the function $\sigma (t, x)$ depends only upon the
variable $\theta = x-t$. This is a theoretically reasonable assumption
as it is the result of assuming that the theory is time translation
invariant. Most of the functions used for $\sigma (t,x)$ in the
literature satisfy this condition. 

The initial forward rate curve $f(t_0, x)$ has to be specified. The
field values of $A(t,x)$ on the rest of the boundary points of the
domain are arbitrary and are integration variables.  The presence of
the second term in the action given in (\ref{sf}) seems to be
justified from the phenomenology of the forward rates \cite{b2} and is
not ruled out by no arbitrage . 

The moment generating functional for the quantum field theory is given by
the Feynman path integral as
\begin{eqnarray}
Z[J]={1\over Z}\int DA e^{\int_{t_0}^{t_*}dt\int_t^{t+T_{FR}}dxJ(t,x)A(t,x)}e^{S[A]}
\end{eqnarray}

On performing the calculation (details are provided in
\cite{baaquie}), we obtain
\begin{equation}
\label{za}
Z[J]=e^{
{1\over2}\int_{t_0}^{t_*}dt\int_0^{T_{FR}}d\theta d\theta'J(t,\theta)D(\theta,\theta';t,T_{FR})J(t,\theta')}
\end{equation}
where $\theta = x-t,\,\, \theta' = x'-t$ and the propagator
$D(\theta,\theta';t,T)$ is given by 
\begin{eqnarray}
&&D(\theta,\theta';t,T_{FR})=\frac{\mu T_{FR}}{
\sinh^3(\mu T_{FR})}\Big[\sinh\mu(T_{FR}-\theta)\sinh\mu\theta'\nonumber\\
&&\{1+\sinh^2(\mu T_{FR})\Theta(\theta-\theta')\}
+\sinh\mu(T_{FR}-\theta')\sinh\mu\theta\nonumber\\
&&\{1+\sinh^2(\mu T_{FR})\Theta(\theta'-\theta)\}+\cosh(\mu
T_{FR})\nonumber\\
&&\{\sinh\mu\theta\sinh\mu\theta'+\sinh\mu(T_{FR}-\theta)\sinh\mu(T_{FR}-\theta')\}\Big]
\end{eqnarray}
The above results are for unconstrained boundary conditions. It is,
however, well known that short term interest rates are heavily
influenced by central banks. Hence, it is reasonable to treat the
field at the boundary where $t=x$ differently. If we assume that the
field at that boundary (i.e. $A(t,t)$) is distributed normally with
variance $a$, we obtain the propagator 
\eq
D_1(\theta, \theta') = D(\theta, \theta') - \frac{D(0, \theta)D(0,
  \theta')}{D(0,0) + a}
\qe
(It can be readily seen that the mean of the field at the boundary
does not affect any of the  results due to the no arbitrage condition.)

To understand the significance of the propagator
$D(\theta,\theta';t,T_{FR})$ we note that the correlator of the field
$A(t,\theta)$, is given by 
\begin{equation}
\label{aa}
E(A(t,\theta)A(t',\theta'))= \delta(t-t')D(\theta,\theta';t,T_{FR})
\end{equation}

It can be readily shown that the no arbitrage condition is satisfied
only when 
\begin{eqnarray}
\label{noara}
\alpha(t,x)&=&\sigma(t,x)\int_{t}^{x}dx' D(x,x';t,T_{FR})\sigma(t,x')
\end{eqnarray}
In the limit $\mu \rightarrow 0$, $D \rightarrow 1$ and we obtain the
well known result 
\begin{equation}
\label{hjmnoar}
\alpha(t,x) = \sigma(t, x)\int_t^x dx' \sigma(t,x')
\end{equation}
for the one factor HJM model.

Following Bouchaud \cite{b2}, we use the daily closing prices for eurodollar
futures prices as a measure of the forward rates. The eurodollar
futures prices are linearly interpolated to calculate the forward
rates at 3 month intervals. The 3 month deposit rate that the
eurodollar futures actually represents is taken to be a good
approximation to the instantaneous forward rate. The data used for
this paper are the same as that used in \cite{b2}. The data cover the
1990s and the length of the dataset is 846 trading days and forward
rates 7 years into the future are available.

We parametrize the forward rates as $f(t, \theta)$ rather than $f(t,
x)$ as this considerably simplifies the analysis considerably since
the domain shape in the $(t, \theta)$ variables is rectangular. 

We concentrate mainly on the following quantities (again partially 
following \cite{b2})
\begin{eqnarray} 
  \label{eq:expsigmae} 
  V (\theta) &=& \sqrt{<\delta f^2(t, \theta)>} \\ 
  \label{eq:expC} 
  C(\theta) &=& \frac{<\delta f(t, \theta_{min}) (\delta f(t, \theta) - \delta f(t,
    \theta_{min}))>}{<\delta f^2(t, \theta_{min})>}\\
  r(\theta) &=& \frac{V(\theta)}{C(\theta)+1}
\end{eqnarray}
with the differences being taken over one trading day ($\epsilon$),
$\delta f(t, \theta) = f(t+\epsilon, \theta) - f(t, \theta)$ and
$\theta_{min}$ being three months. We assume that there are 250
trading days in a year. In the following analysis, we use the
discretization $\delta(0) = \frac{1}{\epsilon}$.

Using the one factor HJM model, we can derive the following
expressions for the above quantities which are accurate to zeroth
order in $\epsilon$
\begin{eqnarray}
\label{eq:hjmsigmae}
V_{HJM} (\theta) &=& \sigma(\theta) \sqrt{\epsilon}\\
\label{eq:hjmC}
C_{HJM}(\theta) &=& \frac{\sigma(\theta)}{\sigma(\theta_{min})}-1\\
\label{eq:hjmr}
r_{HJM}(\theta) &=& \sigma(\theta_{min})\sqrt{\epsilon}
\end{eqnarray}
In deriving this equation, we have discretized the Brownian motion
process W as $W(t) = \sqrt{\frac{1}{\epsilon}}x$ where $x$ is a random
number with the standard normal distribution. We particularly note
that the ratio $r_{HJM}(\theta)$ is {\em independent} of $\sigma(\theta)$
and is in fact constant. The ratio as calculated from the data is
shown in figure \ref{fig:comparison} and can be seen to be far 
from constant. Hence we see that the time translation invariant one
factor HJM model is inconsistent with the real evolution of the FRC
for {\em any} choice of function $\sigma(\theta)$.

Using the unconstrained quantum field theoretic model, we can again
derive the expressions for the above quantities to zeroth order 
accuracy in $\epsilon$ to obtain
\begin{eqnarray}
\label{eq:qftsigmae}
V_{QFT}(\theta) &=& \sigma(\theta)\sqrt{D(\theta, \theta; t, T_{FR})\epsilon}\\
\label{eq:qftC}
C_{QFT}(\theta) &=& \frac{\sigma(\theta)D(\theta, \theta_{min}; t,
  T_{FR})}{\sigma(\theta_{min})D(\theta_{min}, \theta_{min}; t, T_{FR})}-1
\end{eqnarray}

\begin{figure}[t]
  \begin{center}
    \epsfig{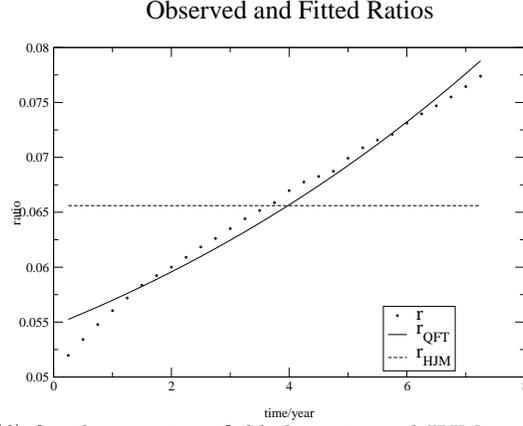}
    \caption{The observed and fitted $r(\theta)$ for the quantum field
      theoretic and HJM models. It can be seen that the quantum field
      theoretic model is in much better agreement with the data.}
    \label{fig:comparison}
  \end{center}
\end{figure}

\begin{table}
  \begin{tabularx}{\linewidth}{|X|X|X|} \hline
    & $\mu (\mathrm{year}^{-1})$ & $\sigma(\theta_{min}) (\mathrm{year}^{-1})$ \\ \hline
    Least Squares & 0.0822 & 0.0308 \\ \hline
    Bootstrap (90\% C.I.)& (0.080, 0.085) & (0.0299, 0.0317)  \\ \hline
    Serial Segments (90\% C.I.)& (0.081, 0.099) & (0.028, 0.033)\\ \hline
  \end{tabularx}
\caption{Results obtained for the unconstrained quantum field
  theoretic model.} 
\label{tab:uncons}
\end{table}

The ratio $r(\theta)$ is thus given in this model by
\eq
\label{eq:qftr}
r_{QFT}(\theta) = \frac{\sigma(\theta_{min})\sqrt{\epsilon D(\theta, \theta; t,
T_{FR})} D(\theta_{min}, \theta_{min}; t, T_{FR})}{D(\theta,
\theta_{min}; t, T_{FR})} 
\end{equation}
which is still independent of $\sigma(\theta)$. However, we note that
the ratio is no longer constant. Using this fact, we can fit the ratio
to find $\mu$ and $\sigma(\theta_{min})$. We took the limit $T_{FR}
\rightarrow \infty$ as required and used the Levenberg-Marquardt
method \cite{lev} to obtain the non-linear least squares fit. The
results are shown in table \ref{tab:uncons}. The confidence
intervals were obtained through the bootstrap method\cite{bootstrap}.
An alternative confidence interval was obtained by dividing the data
into series of 500 days starting from the first day, second day and so
on and so forth. The function $r(\theta)$ was calculated and the
parameters fitted for the resulting 346 data sets. The confidence
interval using these serial data sets are also shown in table
\ref{tab:uncons}.  The fitted ratio together with the observed
values are shown in figure \ref{fig:comparison}.

\begin{figure}[h]
  \begin{center}
    \epsfig{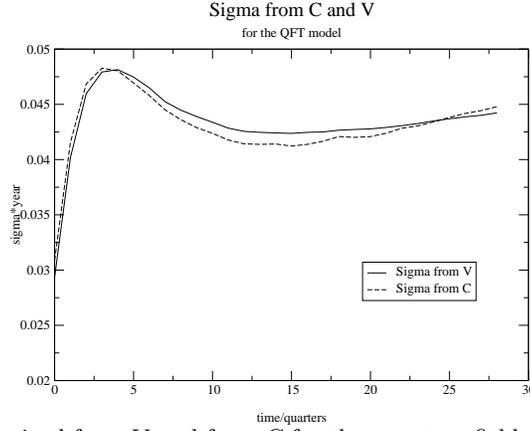}
    \caption{$\sigma(\theta)$ derived from $V$ and from $C$
      for the quantum field theoretic model.}
    \label{fig:sigmas}
  \end{center}
\end{figure}

\begin{figure}[h]
  \begin{center}
    \epsfig{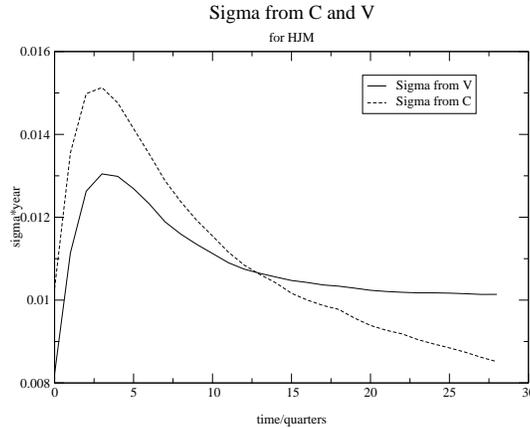}
    \caption{$\sigma(\theta)$ derived from $V$ and from $C$ for
the one factor HJM model}
    \label{fig:hjmsigmas}
  \end{center}
\end{figure}

We can also use equations \ref{eq:qftsigmae} and \ref{eq:qftC} to
obtain two different estimates of the function $\sigma(\theta)$. These
two estimates are plotted in figure \ref{fig:sigmas}. Similar
estimates of $\sigma(\theta)$ for the one factor HJM model are plotted
in figure \ref{fig:hjmsigmas}. As can be readily seen, the HJM model
is seen to be inconsistent with the data. On the other hand, the
quantum field theoretic model is in good agreement with data. It is
also interesting to note that the volatility function for the HJM
model derived from the data is very far from the constant or
exponential forms that are commonly used in the literature.

\begin{figure}[h]
  \begin{center}
    \epsfig{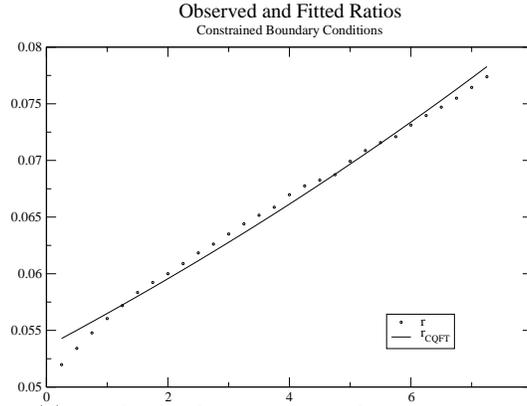}
    \caption{The fit for $r(\theta)$ obtained using the constrained quantum field theoretic model}
    \label{fig:cons_fit}
  \end{center}
\end{figure}

\begin{figure}[h]
  \begin{center}
    \epsfig{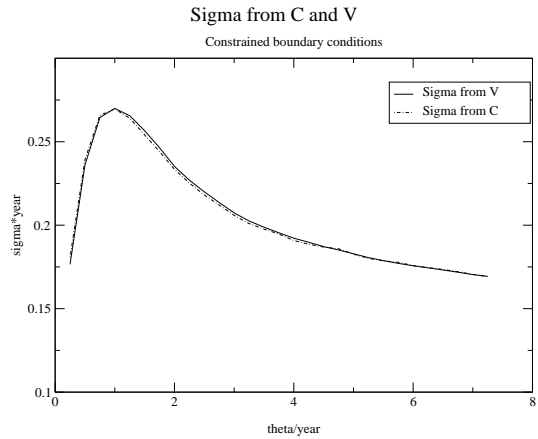}
    \caption{$\sigma(\theta)$ derived from $V$ and from $C$ for
      the constained quantum field theoretic model}
    \label{fig:cons_sigmas}
  \end{center}
\end{figure}

Performing the same procedure for the constrained quantum field
theoretic model, we obtain the results in table \ref{tab:cons}. The
fitted ratio in this case is shown in figure \ref{fig:cons_fit}. The
two estimates of $\sigma(\theta)$ are shown in figure
\ref{fig:cons_sigmas}. The agreement between the two functions is
better than in the case of the unconstrained model as may be expected
due to the additional parameter involved. However, it can be seen from
the large confidence intervals that the model is probably
overspecified since different values of the parameters give rise to
very similar values for $r(\theta)$.

Another quantity that is of great interest is the mean spread between
the forward rates and the spot rate
\eq
\label{eq:exps} 
  s(\theta) = <f(t, \theta) - f(t, \theta_{min})> 
\qe
The spread is the linear sum of two parts : the spread due to the market
price of risk and the spread that results from the no arbitrage
condition in the model. Since we assume that $\sigma$ is only a
function of $\theta$, it follows that $\alpha$ is also only a function
of $\theta$. To calculate the spread due to the model, we assume that
the initial forward rate curve is flat or that the effect of the
initial forward rate curve becomes negligible after a long time.

In that case, the mean spread due to the no arbitrage condition in the
quantum field theoretic model is given by
\begin{equation}
  \label{eq:spread}
  s_{QFT}(\theta) = (\theta-\theta_{min})
\lim_{t\rightarrow\infty}\alpha(t) - \int_{\theta_{min}}^\theta
\alpha(t) dt
\end{equation}
where 
\begin{equation}
  \label{eq:alpha}
  \alpha_{QFT}(t) = \sigma(t) \int_0^t \sigma(\theta) D(t, \theta; t, T_{FR})
  d\theta
\end{equation}

\begin{table}[h]
  \begin{tabularx}{\linewidth}{|X|X|X|X|} \hline
    & $\mu (\mathrm{year}^{-1})$ & $\sigma(\theta_{min}) (\mathrm{year}^{-1})$ &
    $a (\mathrm{year}^{-1})$\\ \hline
    Least Squares & 0.0174 & 0.181 & 0.0024\\ \hline
    Bootstrap (90\% C.I.) & (0.006, 0.020) & (0.158, 0.548) &
    (0.0002, 0.0032)\\
    \hline
    Serial Segments (90\% C.I.) & (0.019, 0.055) & (0.044, 0.203) &
    (0.002, 0.071) \\ \hline
  \end{tabularx}
  \caption{Results obtained for the constrained quantum field theoretic model}
  \label{tab:cons}
\end{table}

\begin{figure}[h]
  \begin{center}
    \epsfig{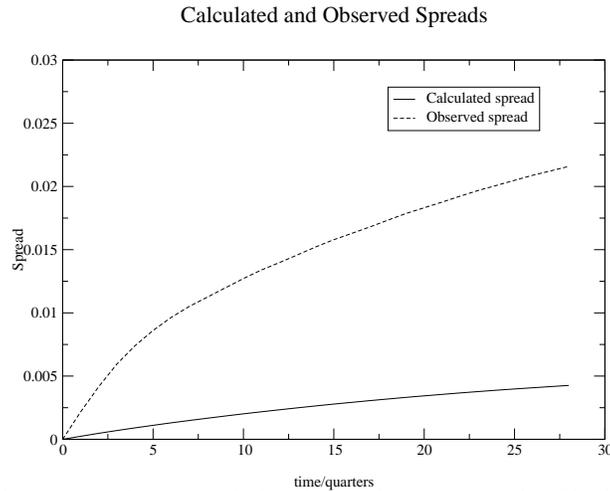}
    \caption{The calculated spread due to no arbitrageand the observed
      mean spreads. The large difference is due to the spread due to
      risk aversion in the market.}
    \label{fig:spread}
  \end{center}
\end{figure}

Using one of the estimates of $\sigma(\theta)$ (using either gives
very similar results), we can calculate the spread due to the no
arbitrage condition by numerical integration. Due to the relative
inaccuracy of the estimation of $\sigma(\theta)$ in the first place, a
trapezoidal integration was considered sufficient. The result together
with the observed spread is shown in figure \ref{fig:spread}. It is
seen that the calculated spread is significantly smaller than the
actual spread which is consistent with the existence of the spread due
to risk aversion. However, we see that a significant portion of the spread
might be derived from the way the forward rate curve evolves. A very
similar result is obtained when the constrained quantum field
theoretic model is used.

To summarize, we have proposed a new way to test the one factor, time
translation invariant Heath-Jarrow-Morton and Baaquie's one factor,
time translation invariant quantum field theoretic model for the
evolution of forward rates using historical eurodollar futures data.
We have found that the one factor HJM model can be rejected while the
quantum field theoretic model is consistent with the data. We
also find that a quantum field theoretic model with constrained
boundary conditions to reflect the special nature of the spot rate is
also consistent with the data but the parameters of the model cannot
be sufficiently accurately derived using this method. We also show
that a significant portion of the spread can be explained by the
quantum field theoretic model.

We would like to thank Jean Phillipe Bouchaud for interesting
discussions and Science and Finance for kindly providing us with the
data.

\end{document}